\documentclass[conference]{IEEEtran} 

\usepackage{graphicx}
\usepackage{euscript,psfrag}
\usepackage[cmex10]{amsmath}
\usepackage{amssymb}                                                        
\usepackage{graphicx}
\usepackage{psfig}
\usepackage{cite}
\usepackage{makeidx}
\usepackage{amsmath}
\usepackage{mathrsfs}
\usepackage{array}
\usepackage{tabularx}
\usepackage{multirow}

\usepackage{url}
\usepackage{hyperref}
\usepackage{tabularx}

\newtheorem{theorem}{Theorem}[section]
\newtheorem{lemma}[theorem]{Lemma}
\newtheorem{definition}[theorem]{Definition}

\newtheorem{remark}[theorem]{Remark}

\newtheorem{problem}[theorem]{Problem}

\usepackage{graphics} 
\usepackage{epsfig} 
\usepackage{mathptmx} 
\usepackage{times} 
\usepackage{amsmath} 
\usepackage{amssymb}  

\usepackage{caption}
\usepackage{subcaption}
\usepackage{dblfloatfix}

\IEEEoverridecommandlockouts                              
\newcommand\beqn{\begin{eqnarray*}}
	\newcommand\eeqn{\end{eqnarray*}}
\newcommand\beqa{\begin{eqnarray}}
\newcommand\eeqa{\end{eqnarray}}

\graphicspath{{./figures/}}

\begin{document}
\title{Privacy-Preserving Aggregation of Controllable	Loads to Compensate Fluctuations in Solar Power}
\author{
	\IEEEauthorblockN{Jin Dong\IEEEauthorrefmark{1},
		Teja Kuruganti\IEEEauthorrefmark{1},
		Seddik Djouadi\IEEEauthorrefmark{2},
		Mohammed Olama\IEEEauthorrefmark{1},
		 and Yaosuo Xue\IEEEauthorrefmark{1}
		}
	\IEEEauthorblockA{\IEEEauthorrefmark{1} Oak Ridge National Laboratory, Oak Ridge, TN 37831}
	\IEEEauthorblockA{\IEEEauthorrefmark{2} Department of Electrical Engineering and Computer Science, University of Tennessee, Knoxville, TN 37996}
	\IEEEauthorblockA{Email: {\tt\small \{dongj, kurugantipv, olamahussemm, xuey\}@ornl.gov},  {\tt\small mdjouadi@utk.edu}}
	\thanks{This manuscript has been authored by UT-Battelle, LLC under Contract No. DE-AC05-00OR22725 with the U.S. Department of Energy. The United States Government retains and the publisher, by accepting the article for publication, acknowledges that the United States Government retains a non-exclusive, paid-up, irrevocable, world-wide license to publish or reproduce the published form of this manuscript, or allow others to do so, for United States Government purposes.
		The Department of Energy will provide public access to these results of federally sponsored research in accordance with the DOE Public Access Plan {http://energy.gov/downloads/doe-public-access-plan}.}}
\maketitle

\begin{abstract}

Cybersecurity and privacy are of the utmost importance for safe, reliable operation of the electric grid. It is well known that the increased connectivity/interoperability between all stakeholders (e.g., utilities, suppliers, and consumers) will enable personal information collection. Significant advanced metering infrastructure (AMI) deployment and demand response (DR) programs across the country, while enable enhanced automation, also generate energy data on individual consumers that can potentially be used for exploiting privacy. Inspired by existing works which consider DR, battery-based perturbation, and differential privacy noise adding, we novelly consider the aggregator (cluster) level privacy issue in the DR framework of solar photovoltaic (PV) generation following.
Different from most of the existing works which mainly rely on the charging/discharging scheduling of rechargeable batteries, we utilize controllable building  loads to serve as virtual storage devices to absorb a large portion of the PV generation while delicately keeping desired noisy terms to satisfy the differential privacy for the raw load profiles at the aggregator level.
This not only ensures differential privacy, but also improves the DR efficiency in load following since part of the noisy signal in solar PV generation has been filtered out.
In particular, a mixed integer quadratic optimization problem is formulated to optimally dispatch a population of on/off controllable loads to achieve this privacy-preserving DR service.

\end{abstract}


\section{INTRODUCTION} \label{sec:introduction}


An important feature of the smart grid is the demand response (DR) mechanism that provides customers with flexibility to meet their energy needs.
Intermittent and volatile production of renewable energy leads to an unavoidable incorporation between customers and energy sources that increases the need for additional balancing resources.
In addition to battery energy storage (BES), virtual storage devices such as thermostatically controlled loads (TCLs) play as an alternative storage to help restore demand-supply balance.
However, significant advanced metering infrastructure (AMI) deployment and DR programs, while enable enhanced automation, also generate energy data on individual consumers that can potentially be used for exploiting privacy \cite{horne2015privacy,zhao2017analysis}.
There exist major concerns about consumers' privacy when aggregators provide DR services using controllable building loads, especially after an intruder gains access to the aggregated power consumption data or the aggregator releases this information to the public.
According to a study by NIST \cite{grid2010introduction}, obtaining near-real-time data regarding energy consumption may infer when and how long a residence or facility is occupied, where people are in the structure, what they are doing, what’s their favorite team, what’s their political affiliation, and so on.

To protect the residential users' privacy from malicious attacks, many secure data aggregation schemes have been proposed in the literature, i.e., encryption mechanism, battery-based perturbations, and noise adding.
Homomorphic encryption was utilized in \cite{lu2012eppa,li2010secure}  to allow a semi-trust gateway to aggregate consumption reports in a specific residential area.
End-to-end encryption is a straightforward way to hide the communication content and preserve users’ privacy, but at the same time heavily increases communication overhead and computational latency for lightweight embedded systems \cite{liu2017achieving}.


An alternative way is to  bring uncertainty into the original raw data by directly adding noise signal, especially using the differential privacy mechanisms \cite{dwork2006calibrating}.
Broadly speaking, differential privacy is a mathematically provable paradigm for privacy-preserving data sharing \cite{mcsherry2007mechanism}.
It ensures that the outputs of neighboring datasets are indistinguishable from randomized noise.
Various approaches based on differential privacy have been introduced in \cite{yang2014optimal,won2016privacy,barbosa2016technique} for smart meter reading and appliance usage using directly noise addition.

Inspired by existing works which consider DR, battery-based perturbation, and differential privacy noise adding under different scenarios, we novelly consider the privacy issue in the DR framework of solar photovoltaic (PV) generation tracking.
Different from most of the existing works that mainly rely on the charging/discharging scheduling of rechargeable batteries, we utilize controllable building loads to serve as virtual storage devices to track a pre-specified aggregated load profile so that the raw power consumption will be distorted and the differential privacy for personal information can be satisfied.
The main objective of our work is to reduce both fluctuations in solar PV generation and potential privacy leakage while providing certain user satisfaction.
To the best of our knowledge, we are the first to study this challenging problem of increasing solar PV penetration level while providing privacy protection.

\begin{figure} [!bt]
	\begin{center}
		\includegraphics[width= 3.5 in]{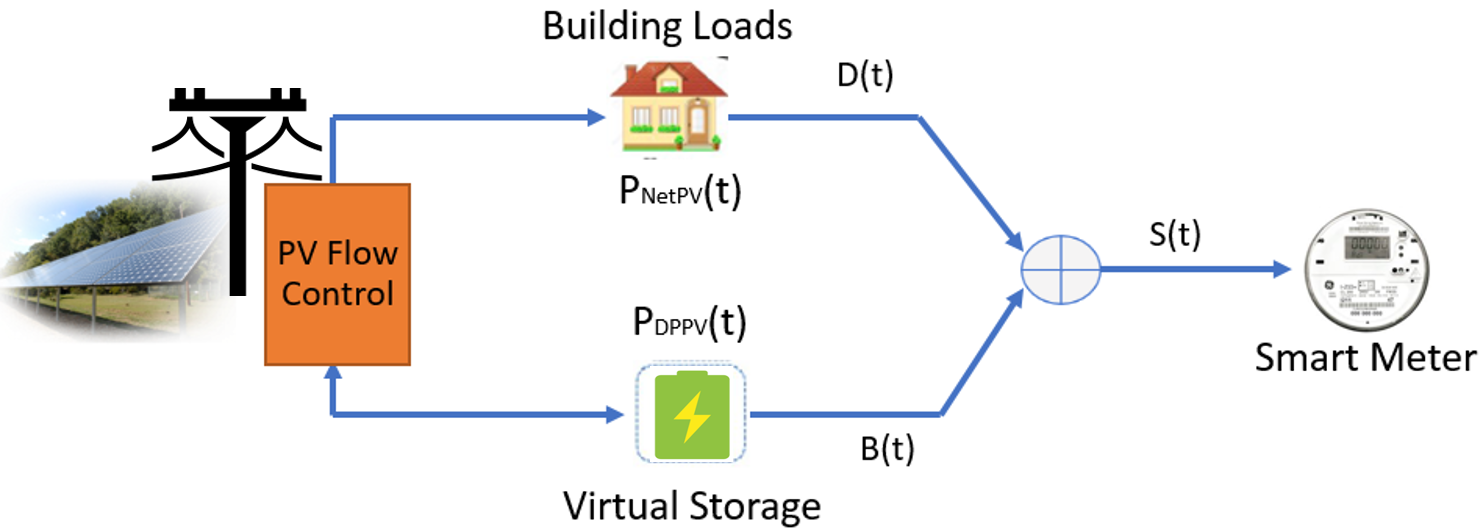}
		\caption{PV-based perturbation mechanism}
		\label{fig:blh2}
	\end{center}
\end{figure}

In this paper, we propose a hybrid method, which is a combination of battery-based perturbation and noise adding.
We denote that, $D(t)$ and $B(t)$ in Fig. \ref{fig:blh2} correspond to base loads and differential privacy signal, respectively.
On the one hand, by filtering out the high frequency components in PV generation, it relieves the pressure of generation following for slow-responsive controllable loads.  
On the other hand, by smartly picking the mid-high frequency components, desired differential privacy noise signal $B(t)$ is perfectly computed, and then implemented through virtual storage devices.
The main privacy concern here is to ensure two aggregated datasets are indistinguishable by delicately keeping the inherent noisy signal inside local solar PV generation.
At last, the performance of the proposed design is verified through controllable Heating, Ventilation and Air-Conditioning (HVAC) loads based on the DR control technique introduced in \cite{jin2017adaptive,dong2018model}.

The remainder of this paper is organized as follows.
Section \ref{sec:bfmodel} introduces the background for differential privacy and problem formulation.
Section \ref{sec:main} then derives the optimization formulation for aggregated HVAC loads.
Section \ref{sec:simulation} presents the simulation results to validate the tracking performance using a set of real temperature and solar power data.
Finally, Section \ref{sec:conclusion} summarizes the paper and presents the conclusions.

\section{Preliminary and problem formulation}\label{sec:bfmodel}

In this section, we will discuss the basic principle for differential privacy and the building thermal model.
By adding random noise, typically with Laplacian \cite{le2014differentially} distribution, to the measurement data submitted by the customers, the differential private aggregation scheme will not increase system complexity.
On the one hand, this contributes to reducing the capital and maintenance cost by purchasing smaller size battery storage.
On the other hand, considering the intermittent renewable resources, this privacy-in-the-loop strategy will contribute to absorbing the PV high frequency uncertain fluctuations and, therefore, reducing stress to the grid.

\subsection{Differential Privacy}

Broadly speaking, differential privacy is a privacy-preserving mechanism, which is independent of the background knowledge of the adversary \cite{mcsherry2007mechanism,dwork2008differential}.   
It guarantees that the probabilities that two neighboring data sets that have the same output are quite close so that the probability dilation is bounded by $ exp(\epsilon) $.
By doing so, the adversary can hardly infer a single data record by manipulating outputs. The definition and detailed description of the standard $ \epsilon $-differential privacy can be found in \cite{dwork2008differential}.

Mathematically, $ \mathcal{M} $ satisfies $ \epsilon $-differential privacy, if for all possible data sets, any individual (say, Alex), and all possible result set $\mathbb{R} \subseteq $ Range($ \mathcal{M} $):
\beqa
\frac{Pr(\mathcal{M}(\text{DB with Alex}) \in \mathbb{R})}{Pr(\mathcal{M}(\text{DB without Alex}) \in \mathbb{R})} \leq e^\epsilon \approx 1\pm \epsilon,
\eeqa
where DB is short for Database.
Note that $ \epsilon $ is denoted as the privacy budget and specifies the privacy degree. Intuitively, a lower $ \epsilon $ means a better privacy and typically $\epsilon = 0.1$.

\begin{definition}
	Given an $n$-dimension dataset $D^n$, a randomized algorithm to answer query $A$ is $(\delta,\epsilon)$-differentially private if $\forall x,y \in D^n$ that differs only in one element and all $S \in range(A)$,
	\beqa
	Pr\left[ A(x) \in S \right] \leq e^\epsilon \times Pr\left[ A(y) \in S \right] + \delta,
	\eeqa
	where $range(A)$ denotes the output range of $A$.
\end{definition}

After reviewing differential privacy, a Baseline Aggregation scheme can be drawn to provide differential privacy for real-time aggregated data streams. The basic principle is to add independent Laplace noise into the aggregation result at each time slot \cite{dwork2006calibrating}, aiming at providing $ \epsilon $-differential privacy.

\subsection{Problem Formulation}

In an aggregated control system, the power consumption data is transmitted to the controller.
It is necessary that the information preserves a certain level of privacy such that an adversary or another curious aggregator cannot infer more information about the system states than what is intended.

The overall objective is to utilize the aggregated HVAC loads to compensate fluctuations in solar PV power as well as protecting the users' privacy without jeopardizing thermal comfort.
In other word, we will utilize controllable loads to maximize the locally absorbed PV generation while providing adequate differential privacy covering noise signal.
Hence, we need first to compute the minimum differential privacy noise signal corresponding to the specific solar PV generation, then ensure the controllable loads consume exactly desired amounts of energy (as illustrated by $P_{NetPV}(t)$ in Fig. \ref{fig:blh2}) through DR algorithms.

\begin{problem}
	(Privacy-preserving generation following problem). Given the dataset $D(t)$ and the query function $Q(\cdot)$ over $D(t)$ which returns $P(t)$ (i.e., $Q(D(t)) = P(t)$), we aim to devise a privacy-preserving algorithm $\mathcal
	A$ which extracts $B(t)$ (from local PV generation $ L(t) $) as  noise to the query result to hide $L(t)$ from  adversaries such that $(\delta,\epsilon)$-differential privacy is guaranteed with good $\delta,\epsilon$.
\end{problem}

\section{Privacy-preserving Generation Following}\label{sec:main}

This section discusses the noise construction technique to provide sufficient differential privacy. It is followed by the control design, which relies on model predictive control (MPC) scheme to regulate the aggregated power consumption of a population of HVAC loads to track a desired profile.

\subsection{Differential Privacy Noise Calculation}

In this subsection, to derive the necessary noise signal for differential privacy, we depart from exploring the definition of $ \left(\epsilon,\delta \right) $-differential privacy.

In the original derivation for differential privacy \cite{dwork2006calibrating}, Dwork presented a general protocol to implement differential privacy utilizing the concept of \textit{global sensitivity} $\Delta Q$.
Given a query set $ Q \in \mathbb{Q} $, the global sensitivity $\Delta Q$ is the maximal $\mathcal{L}_1$ distance (difference) between the exact query results on any two neighboring datasets $D_1$ and $D_2$, i.e.
\beqa
\Delta Q &=& \max_{D_1,D_2}\| Q(D_1), Q(D_2) \|_1 \nonumber \\
&=& \max_{D_1,D_2}|Q(D_1)-Q(D_2)|,
\eeqa
for all $D_1, D_2$ that differ in at most one element.

Informally, the sensitivity $ \Delta Q $ is the maximum amount the result $ Q(\cdot) $ can change, given any change to a single users' data.
\begin{lemma}
	When $\Delta Q = 1$, the function $Q(\cdot) $ achieves $ \left(\epsilon,\delta \right) $-differential privacy if noise from the Laplacian distribution is added to it \cite{dwork2006calibrating}.
\end{lemma}

It should be mentioned that, although we focus on the Laplace noise in this paper, our approach is easily extendable to other  mechanisms that satisfy requirements by differential privacy.

\begin{remark}
	The $\Delta Q$ usually depends on the data domain $\mathbb{D}$ and the query set $Q$, but not the actual data.
	Therefore, we simply assume such a constant $\Delta Q$ is public knowledge to everyone, including the adversary.
\end{remark}

The Laplace Mechanism, $M_L$, outputs a randomized result $R$ on dataset $D$, following a Laplace distribution with mean $Q(D)$ and magnitude $\frac{\Delta Q}{\epsilon}$, i.e.,
\beqa
Pr(M_L(Q,D)=R)\propto exp\left( \frac{\epsilon}{\Delta Q}\| R-Q(D) \|_1 \right).  \label{eq:lap1}
\eeqa

This is equivalent to adding $m$-dimensional independent Laplace noise to each query in $Q$, that is, $M(Q,D) = Q(D)+Lap(\frac{\Delta Q}{\epsilon})^m $, in which $Lap(\frac{\Delta Q}{\epsilon})$ is a random variable following a zero-mean Laplace distribution with scale $\frac{\Delta Q}{\epsilon}$.
The Laplace noise is drawn from Laplace distribution with probability density function (PDF)
\beqa
f(x) = \frac{1}{2\lambda} exp(-\frac{|x|}{\lambda}),  \label{eq:lapnoise}
\eeqa
where the variance of $Lap(\lambda)$ is $2 \lambda^2 = \frac{2 \Delta Q^2}{\epsilon^2}$.
Since the Laplace noise injected into each of the $m$ query result is independent, the overall expected squared error of the query answers obtained by the Laplace mechanism is $ \frac{2m \Delta Q^2}{\epsilon^2}$ \cite{yuan2015optimizing}.
\begin{remark}
	Note that the amount of error only depends on the sensitivity of the queries, regardless of the records in dataset $D$. Therefore, we can obtain the desired noise signal to ensure differential privacy (denoted as $ P_{DPPV}(k) $) by sampling through $ f(x) $ (defined in \eqref{eq:lapnoise} and \eqref{eq:lap1}), where $k$ represents time step.
\end{remark}

\begin{remark}
	$ P_{DPPV}(k) $ corresponds to the desired noise signal $B(t)$ in Fig. \ref{fig:blh2}.
\end{remark}

After extracting this noise part from the total PV generation, we can easily compute the remaining net PV generation $ P_{NetPV}(k) $ that needs to be locally consumed by aggregated controllable loads.
$ P_{NetPV}(k) $, which will play as a new reference profile to be followed by the DR program, is defined as follows.
\beqa
P_{NetPV}(k) = P_{PV}(k) - P_{DPPV}(k).
\eeqa

\subsection{DR for $ P_{DPPV} $}
Due to the partially uncontrollable and stochastic nature of renewable resources, the variance of net load of distribution systems increases \cite{yichen}. In effect, TCLs, such as HVAC units, have been used for virtual energy storage.
Hence, the designed privacy noise signal $P_{DPPV}$ can be provided by such virtual storage devices.
The feasibility of this approach has been well recognized in the literature, for example in \cite{brooks2017virtual,lin2017ancillary,jin2017adaptive,dong2018model}.

\subsection{DR for $ P_{NetPV} $} \label{sec:MPC}

This section develops a centralized MPC-based DR strategy to locally absorb the net PV generation - $P_{NetPV}$.

\subsubsection{Building Thermal Model}
In this section, we describe the typical one-dimensional model used in this work and formulate the control strategy.
The system model was widely utilized in the literature such as \cite{mathieu2013state,dong2018model,dong2018occupancy}.

If we denote the state vector $ X = \left[q   \right], $  input $ U = \left[ Mode_{HVAC} \right]$ (HVAC ON/OFF status), and disturbances $ V =  \left[ T_{out}, Q_{out} \right]$ (outdoor temperature and solar irradiance), we can rewrite the building thermal model into the state-space form as:
\beqa
\dot{X} = A X + B U + G V.  \label{eq:state}
\eeqa
State-space matrices $ A, B, G $ can be obtained for any given building, and disturbance $V$ is recorded for that specific location.

We consider the problem where indoor temperature $ x = q $ is required to remain within certain bounds of a constant dead-band in the presence of the disturbance vector $ V $.
Moreover, we make the indoor temperature $ x_j $ ($j$ corresponds to $j_{th}$ building) track a pre-assigned reference temperature set-points by minimizing the state error $ e_j(k): = \,x_j(k) - {X_r}$ at each time step $ k $, where $ X_r $ is the temperature set-point.

MPC is implemented as a discretized building thermal model with sampling period $\triangle T = 10 $ minute,  ${t_i} = i\Delta T$ which yield the discrete-time model
\beqa
x_{k+1} = A_k x_k + B_k u_k +G_k v_k.  \label{eq:dis}
\eeqa

After ensuring the differential privacy noise signal, we need to design optimal control signals to orchestrate the aggregate demand of HVAC systems follow the net PV power generation $ P_{NetPV} $ without violating the temperature comfort requirement.
\beqa
z(k) := \sum_{j=1}^{N_s} u_j (k) \approx P_{NetPV}(k), \label{eq:usum}
\eeqa
where $N_s$ denotes total number of HVAC units and $ u_j(k) $  denotes the control action taken for the $ j_{th} $  HVAC unit at the $ k_{th} $  time interval.


\subsubsection{Cost Functions}
We formulate the control strategy as an optimization problem, whose objective consists of two parts.
The first part describes the performance of privacy protection while the second part characterizes the user comfort satisfaction ($e_i(k)$ denotes the temperature deviation from temperature setpoint).
\beqa
J &=& \sum_{k=1}^{N_p} \Big\{ Q(z(k) - P_{NetPV}(k))^2 + R(\sum_{i=1}^{N_s}  e_i(k)^2) \Big\} ,  \label{eq:qpcost}
\eeqa
where $ N_p $ represents the prediction horizon with $ Q $ and $ R $ being weighting factors.

\subsubsection{Constraints}
In the MPC problem, we have both states and control input constraints.
For temperature state constraint, we set $ x \in \left[22.5^ \circ C, 23.5^ \circ C \right] $.
And for the discrete ON/OFF control signal, the binary inputs $ u \in \lbrace 0, \, 1 \rbrace $.
Notice here, ``0" means HVAC OFF, while ``1" means HVAC ON.

\section{Performance Evaluation} \label{sec:simulation}

In this section, we evaluate the proposed privacy-preserving aggregation algorithm through a numerical example.
We consider a central coordinator that collects total PV generation, computes desired differential privacy noise signal, and allocates energy to a population of HVAC loads to {minimize difference between total power consumption and net PV generation}.

Based on the MPC design introduced in Sec. \ref{sec:MPC}, we utilize $ 100 $ buildings, each of which is equipped with identical building thermal model \eqref{eq:dis}.
It is worth mentioning that the simulation time is 432 time steps, which means $ 10 $ mins per time step for three days.
Both PV generation and weather profiles are picked for these days from a local station.

\subsection{Differential Privacy Noise}
\begin{figure} [!hbt]
	\begin{center}
		\includegraphics[width= 3 in]{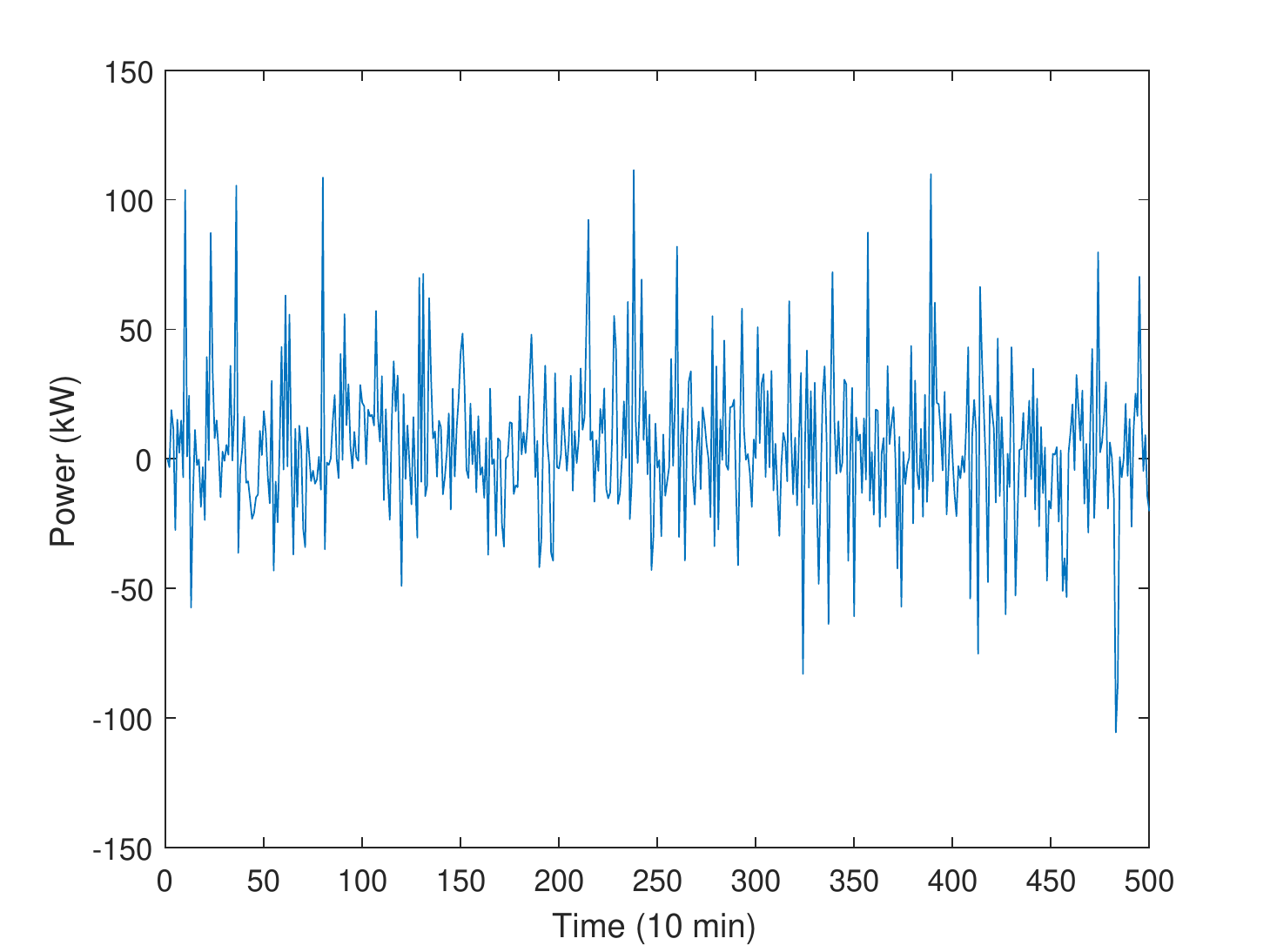}
		\caption{Desired Laplacian noise}
		\label{fig:lapnoise}
	\end{center}
\end{figure}

\begin{figure} [!hbt]
	\begin{center}
		\includegraphics[width= 3 in]{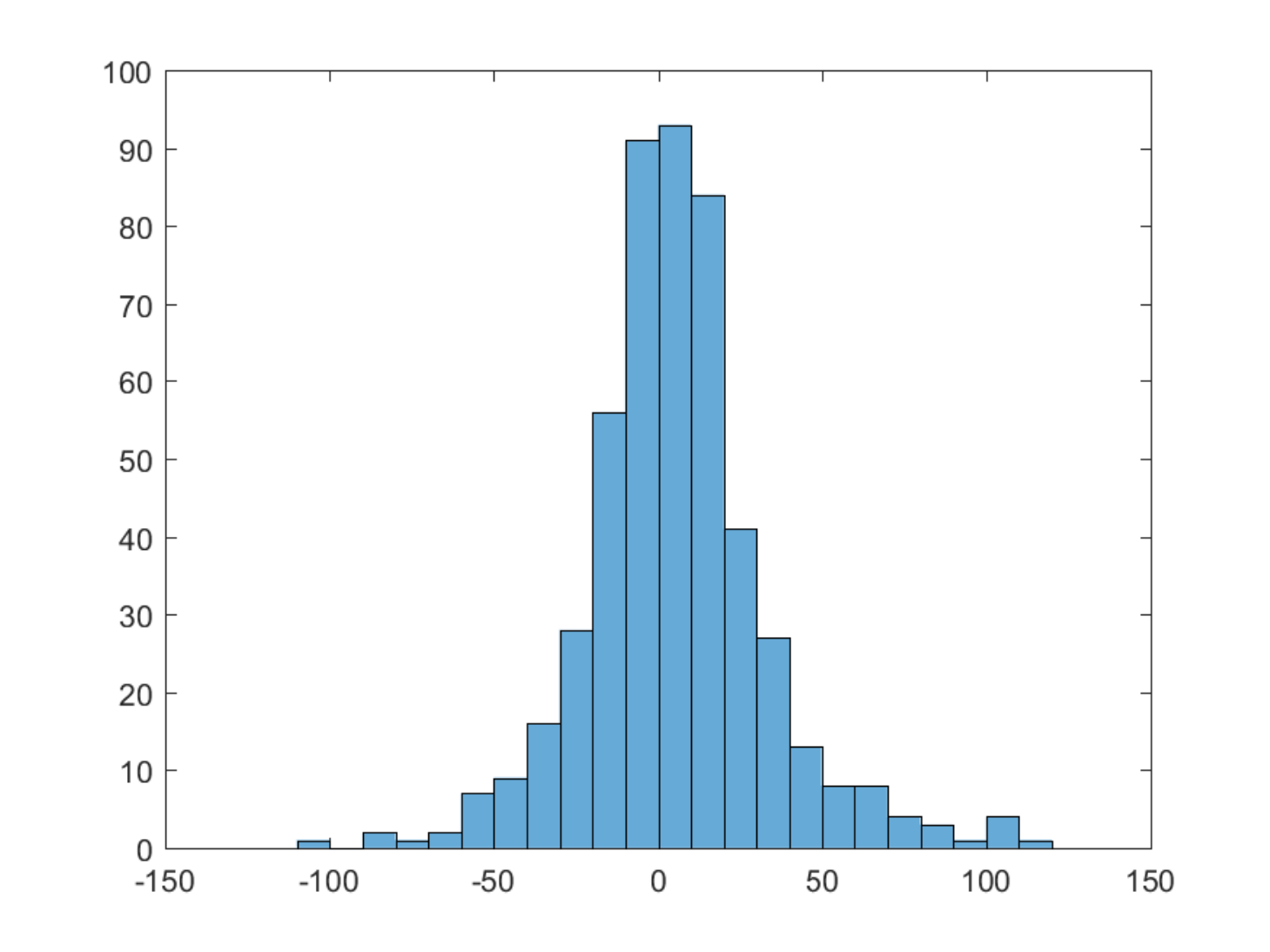}
		\caption{Histogram of the Laplacian noise}
		\label{fig:histogram}
	\end{center}
\end{figure}

\begin{figure} [!h]
	\begin{center}
		\includegraphics[width= 2.8 in]{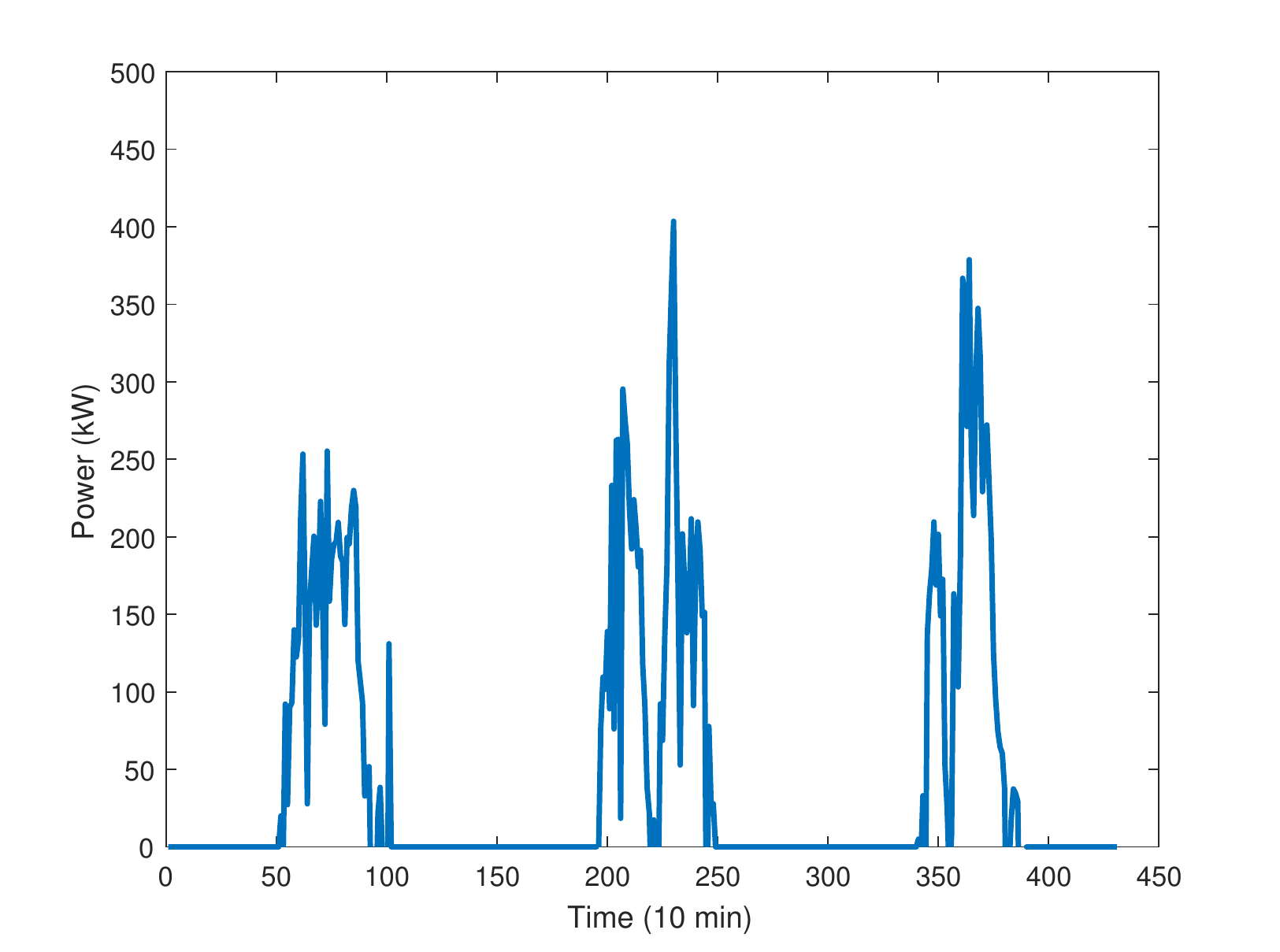}
		\caption{Net reference PV power generation ($ P_{NetPV} $)}
		\label{fig:netPV}
	\end{center}
\end{figure}

The desired noise is generated by $Lap(\mu,b)$, with $\mu = 0$ and $b = \frac{\Delta Q}{\epsilon}$ (corresponding to the raw PV generation measurement), and $\epsilon = 0.1$.
Figs. \ref{fig:lapnoise} and \ref{fig:histogram}, respectively, illustrate the detailed noise signal for each time step and the overall histogram for the generated noise.
From Fig. \ref{fig:histogram}, we observe that Laplace noise is obtained.
Then, the net PV generation $ P_{NetPV}(k) $ profile is depicted in Fig. \ref{fig:netPV}.

\subsection{Load Tracking Performance for $ P_{NetPV} $}

\begin{figure} [!h]
	\begin{center}
		\includegraphics[width= 2.8 in]{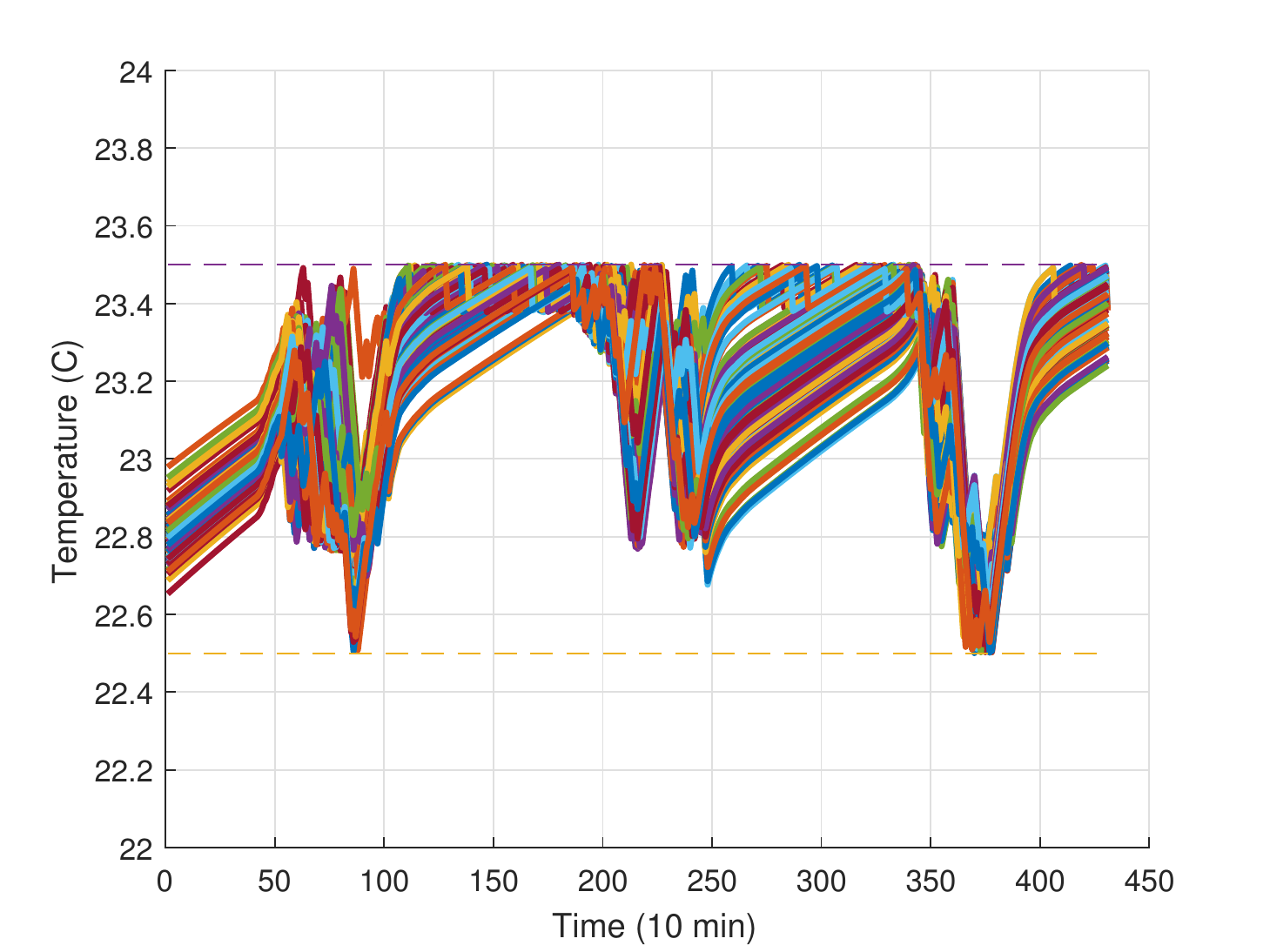}
		\caption{Indoor temperature for 100 buildings when tracking $ P_{NetPV} $ using MPC}
		\label{fig:IndoorTempMPC}
	\end{center}
\end{figure}
\begin{figure} [!h]
	\begin{center}
		\includegraphics[width= 2.8 in]{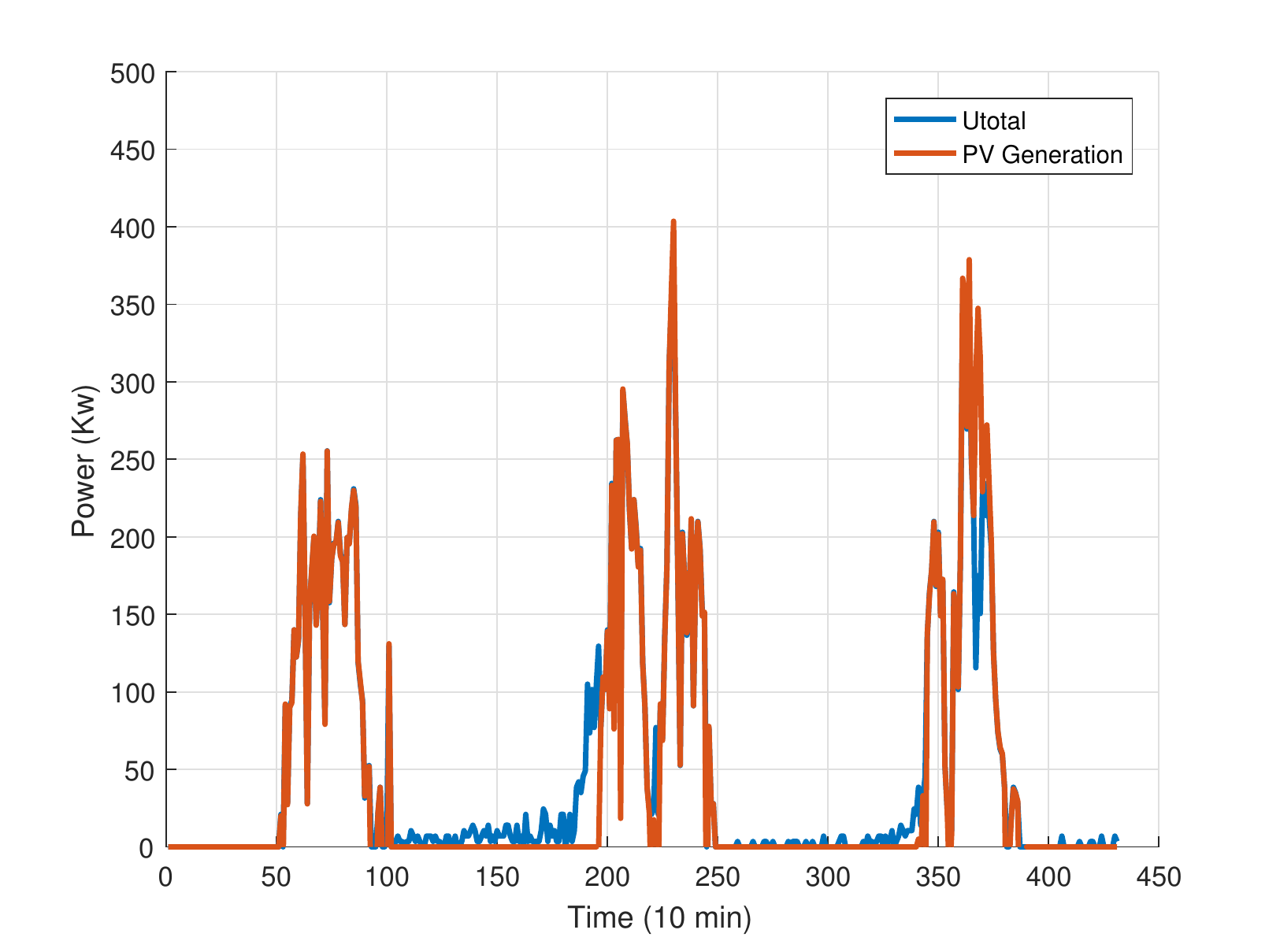}
		\caption{Tracking the $ P_{NetPV} $ signal using MPC control}
		\label{fig:TrackMPC}
	\end{center}
\end{figure}

Based on the net PV generation profile, we present simulation results  for the developed MPC strategy of all 100 buildings in Figs. \ref{fig:IndoorTempMPC} -  \ref{fig:TrackMPC}.
The indoor temperatures  are plotted in Fig. \ref{fig:IndoorTempMPC}.
We can observe that the indoor temperatures are strictly bounded by the preassigned comfort band, utilizing discrete ON/OFF control signals.
After checking the temperatures in bound, we need to evaluate the tracking performance depicted in Fig. \ref{fig:TrackMPC}.
It can be seen from Fig. \ref{fig:TrackMPC} that satisfied tracking performance has been achieved, in particular with the presence of highly dynamic solar PV generation after extracting the designed noise signal.
Therefore, we claim that these 100 HVAC loads are well coordinated to track the PV generation using the developed optimal control strategy.


\section{CONCLUSIONS AND FUTURE WORKS} \label{sec:conclusion}

In this paper, the privacy issues associated with DR that aim to improve security and  efficiency of aggregated controllable loads for mitigating fluctuations in solar PV generation are addressed.
Different from most of the existing works which mainly rely on the charging/discharging scheduling of rechargeable batteries, we utilize controllable building loads to serve as virtual storage devices to track a pre-specified aggregated load profile so that the raw power consumption will be distorted and the differential privacy for personal information can be guaranteed.
A differential privacy based aggregation algorithm is proposed to compensate for fluctuations in solar power as well as to protect the users' privacy without jeopardizing user comfort.
Considering the intermittent renewable resources, this privacy-in-the-loop strategy will contribute to filter out high frequency uncertain fluctuations into the grid.
Optimal coordination of controllable loads is then investigated to simultaneously track both the load and noise profiles.
A numerical example is provided to optimally dispatch an aggregate of on/off HVAC units to track the designed net PV generation signal after filtering out the privacy signal.
It's worth mentioning that, this framework can be generalized to any controllable loads.

In our future work, a distributed cooperative optimization-based control algorithm will be combined with this framework to address the large-scale HVAC system control problem.
Another interesting direction is to design a more complex privacy-preserving DR framework, whose objective is to ensure the privacy of each individual user under the aggregator.

\bibliographystyle{IEEEtran}
\bibliography{DJ_draft,Ref_LF}  

\begin{thebibliography}{10}
\providecommand{\url}[1]{#1}
\csname url@samestyle\endcsname
\providecommand{\newblock}{\relax}
\providecommand{\bibinfo}[2]{#2}
\providecommand{\BIBentrySTDinterwordspacing}{\spaceskip=0pt\relax}
\providecommand{\BIBentryALTinterwordstretchfactor}{4}
\providecommand{\BIBentryALTinterwordspacing}{\spaceskip=\fontdimen2\font plus
\BIBentryALTinterwordstretchfactor\fontdimen3\font minus
  \fontdimen4\font\relax}
\providecommand{\BIBforeignlanguage}[2]{{%
\expandafter\ifx\csname l@#1\endcsname\relax
\typeout{** WARNING: IEEEtran.bst: No hyphenation pattern has been}%
\typeout{** loaded for the language `#1'. Using the pattern for}%
\typeout{** the default language instead.}%
\else
\language=\csname l@#1\endcsname
\fi
#2}}
\providecommand{\BIBdecl}{\relax}
\BIBdecl

\bibitem{horne2015privacy}
C.~Horne, B.~Darras, E.~Bean, A.~Srivastava, and S.~Frickel, ``Privacy,
  technology, and norms: The case of smart meters,'' \emph{Social science
  research}, vol.~51, pp. 64--76, 2015.

\bibitem{zhao2017analysis}
C.~Zhao, J.~He, P.~Cheng, and J.~Chen, ``Analysis of consensus-based
  distributed economic dispatch under stealthy attacks,'' \emph{IEEE
  Transactions on Industrial Electronics}, vol.~64, no.~6, pp. 5107--5117,
  2017.

\bibitem{grid2010introduction}
N.~S. Grid, ``Introduction to nistir 7628 guidelines for smart grid cyber
  security,'' \emph{Guideline, Sep}, 2010.

\bibitem{lu2012eppa}
R.~Lu, X.~Liang, X.~Li, X.~Lin, and X.~Shen, ``Eppa: An efficient and
  privacy-preserving aggregation scheme for secure smart grid communications,''
  \emph{IEEE Transactions on Parallel and Distributed Systems}, vol.~23, no.~9,
  pp. 1621--1631, 2012.

\bibitem{li2010secure}
F.~Li, B.~Luo, and P.~Liu, ``Secure information aggregation for smart grids
  using homomorphic encryption,'' in \emph{Smart Grid Communications
  (SmartGridComm), 2010 First IEEE International Conference on}.\hskip 1em plus
  0.5em minus 0.4em\relax IEEE, 2010, pp. 327--332.

\bibitem{liu2017achieving}
E.~Liu and P.~Cheng, ``Achieving privacy protection using distributed load
  scheduling: A randomized approach,'' \emph{IEEE Transactions on Smart Grid},
  2017.

\bibitem{dwork2006calibrating}
C.~Dwork, F.~McSherry, K.~Nissim, and A.~Smith, ``Calibrating noise to
  sensitivity in private data analysis,'' in \emph{TCC}, vol. 3876.\hskip 1em
  plus 0.5em minus 0.4em\relax Springer, 2006, pp. 265--284.

\bibitem{mcsherry2007mechanism}
F.~McSherry and K.~Talwar, ``Mechanism design via differential privacy,'' in
  \emph{Foundations of Computer Science, 2007. FOCS'07. 48th Annual IEEE
  Symposium on}.\hskip 1em plus 0.5em minus 0.4em\relax IEEE, 2007, pp.
  94--103.

\bibitem{yang2014optimal}
L.~Yang, X.~Chen, J.~Zhang, and H.~V. Poor, ``Optimal privacy-preserving energy
  management for smart meters,'' in \emph{INFOCOM, 2014 Proceedings
  IEEE}.\hskip 1em plus 0.5em minus 0.4em\relax IEEE, 2014, pp. 513--521.

\bibitem{won2016privacy}
J.~Won, C.~Y. Ma, D.~K. Yau, and N.~S. Rao, ``Privacy-assured aggregation
  protocol for smart metering: A proactive fault-tolerant approach,''
  \emph{Biological Cybernetics}, vol.~24, no.~3, pp. 1661--1674, 2016.

\bibitem{barbosa2016technique}
P.~Barbosa, A.~Brito, and H.~Almeida, ``A technique to provide differential
  privacy for appliance usage in smart metering,'' \emph{Information Sciences},
  vol. 370, pp. 355--367, 2016.

\bibitem{jin2017adaptive}
J.~Dong, M.~Olama, T.~Kuruganti, J.~Nutaro, Y.~Xue, I.~Sharma, and S.~M.
  Djouadi, ``Adaptive building load control to enable high penetration of solar
  {PV} generation,'' in \emph{2017 IEEE Power \& Energy Society General
  Meeting}, 2017.

\bibitem{dong2018model}
J.~Dong, M.~Olama, T.~Kuruganti, J.~Nutaro, C.~Winstead, Y.~Xue, and A.~Melin,
  ``Model predictive control of building on/off {HVAC} systems to compensate
  fluctuations in solar power generation,'' in \emph{2018 9th IEEE
  International Symposium on Power Electronics for Distributed Generation
  Systems (PEDG)}.\hskip 1em plus 0.5em minus 0.4em\relax IEEE, 2018, pp. 1--5.

\bibitem{le2014differentially}
J.~Le~Ny and G.~J. Pappas, ``Differentially private filtering,'' \emph{IEEE
  Transactions on Automatic Control}, vol.~59, no.~2, pp. 341--354, 2014.

\bibitem{dwork2008differential}
C.~Dwork, ``Differential privacy: A survey of results,'' in \emph{International
  Conference on Theory and Applications of Models of Computation}.\hskip 1em
  plus 0.5em minus 0.4em\relax Springer, 2008, pp. 1--19.

\bibitem{yuan2015optimizing}
G.~Yuan, Z.~Zhang, M.~Winslett, X.~Xiao, Y.~Yang, and Z.~Hao, ``Optimizing
  batch linear queries under exact and approximate differential privacy,''
  \emph{ACM Transactions on Database Systems (TODS)}, vol.~40, no.~2, p.~11,
  2015.

\bibitem{yichen}
Y.~Zhang, A.~Melin, M.~Olama, S.~Djouadi, J.~Dong, and K.~Tomsovic, ``Battery
  energy storage scheduling for optimal load variance minimization,'' in
  \emph{2018 IEEE Power Energy Society Innovative Smart Grid Technologies
  Conference (ISGT)}, Feb 2018, pp. 1--5.

\bibitem{brooks2017virtual}
J.~Brooks and P.~Barooah, ``Virtual energy storage through decentralized load
  control with quality of service bounds,'' in \emph{Proc. of American Control
  Conference (ACC)}.\hskip 1em plus 0.5em minus 0.4em\relax IEEE, 2017, pp.
  735--740.

\bibitem{lin2017ancillary}
Y.~Lin, P.~Barooah, and J.~L. Mathieu, ``Ancillary services through demand
  scheduling and control of commercial buildings,'' \emph{IEEE Transactions on
  Power Systems}, vol.~32, no.~1, pp. 186--197, 2017.

\bibitem{mathieu2013state}
J.~L. Mathieu, S.~Koch, and D.~S. Callaway, ``State estimation and control of
  electric loads to manage real-time energy imbalance,'' \emph{IEEE
  Transactions on Power Systems}, vol.~28, no.~1, pp. 430--440, 2013.

\bibitem{dong2018occupancy}
J.~Dong, C.~Winstead, J.~Nutaro, and T.~Kuruganti, ``Occupancy-based {HVAC}
  control with short-term occupancy prediction algorithms for energy-efficient
  buildings,'' \emph{Energies}, vol.~11, no.~9, p. 2427, 2018.

\end{thebibliography}
\end{document}